\documentclass[10]{article}
\usepackage{graphicx}
\newcommand{\beq}{\begin{equation}}
\newcommand{\eeq}{\end{equation}}
\begin{document}
\author{Jean Kaplan, Jacques Delabrouille\\PCC, Coll{\`e}ge de France, Paris}
\title{Some sources of systematic errors on CMB polarized
measurements with bolometers}
\maketitle
\begin{abstract}
Some sources of systematic errors, specific to polarized CMB measurements
using bolometers, are examined. Although the evaluations we show have
been made in the context of the Planck mission (and more specifically the
Planck HFI), many of our conclusions are valid for other experiments
as well.
\end{abstract}
\section{The specifics of CMB polarization signals}
CMB fluctuations are difficult to measure because of their extreme
weakness. Systematics must therefore be very well controlled. This is even
more true for polarization anisotropies, which are expected to be less
than 10\% of the temperature fluctuations.\\
Low frequency noise is a source of troubles for polarization 
as well as for temperature measurements. However, as the polarized signal
depends on the direction of the detector projected in the sky, its
suppression, or ``destriping'' requires a specific treatment. The way
in which polarized destriping can be implemented for Planck is
outlined in the next section.\\
A few definitions useful to the following discussion are given in the 
third section.\\
Polarizers are never perfect: the unwanted
polarization is never totally suppressed and the polarizer direction is
never perfectly known. The impact of these uncertainties is evaluated
in section four.\\
In the fifth section, we  discuss the crucial difficulties linked to
signal differences: calibration, pointing and beam mismatches
between different detectors.\\
We  then consider the question of avoiding elliptical error boxes on
the Stokes parameters, which might be confused with a polarization signal 
when the signal to noise ratio is small. \\
A few concluding remarks are given in the last section. 

\section{polarized destriping}
This section is devoted to the elimination of low frequency noises in the
framework of Planck. It relies on the Planck scanning strategy, which
goes as follows:
The telescope beam rotates 60 times 
around a fixed axis with an opening angle around 85$^\circ$. Then the 
axis is shifted by a few arc-minutes and the beam is again rotated 
60 times etc... 
\begin{figure}[h]  
\begin{center}
    {\includegraphics[scale=0.35]{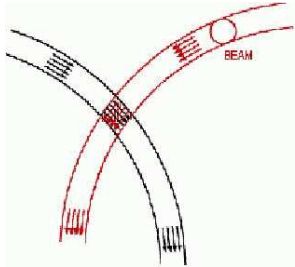}}    
    \caption{Circle crossing} 
    \label{fig:1}  
\end{center}
    \end{figure}   
Averaging over the 60 scans of the same circle suppresses most noise 
with frequencies smaller than the spinning frequency. The remaining 
noise can be described by 1 offset per ring for each of the 3 Stokes parameters,
irrespective of the number of polarized detectors. The circles have many
intersections with each other in the sky.  
 At these circle crossings (see figure \ref{fig:1}), one can use 
the redundancy by asking that the sky Stokes parameters be the same 
along both circles. This results in solving a linear system with the 
Stokes parameter offsets as variable.  
\begin{figure}[!h]
\begin{center}
{\includegraphics[scale=0.4,clip=true]{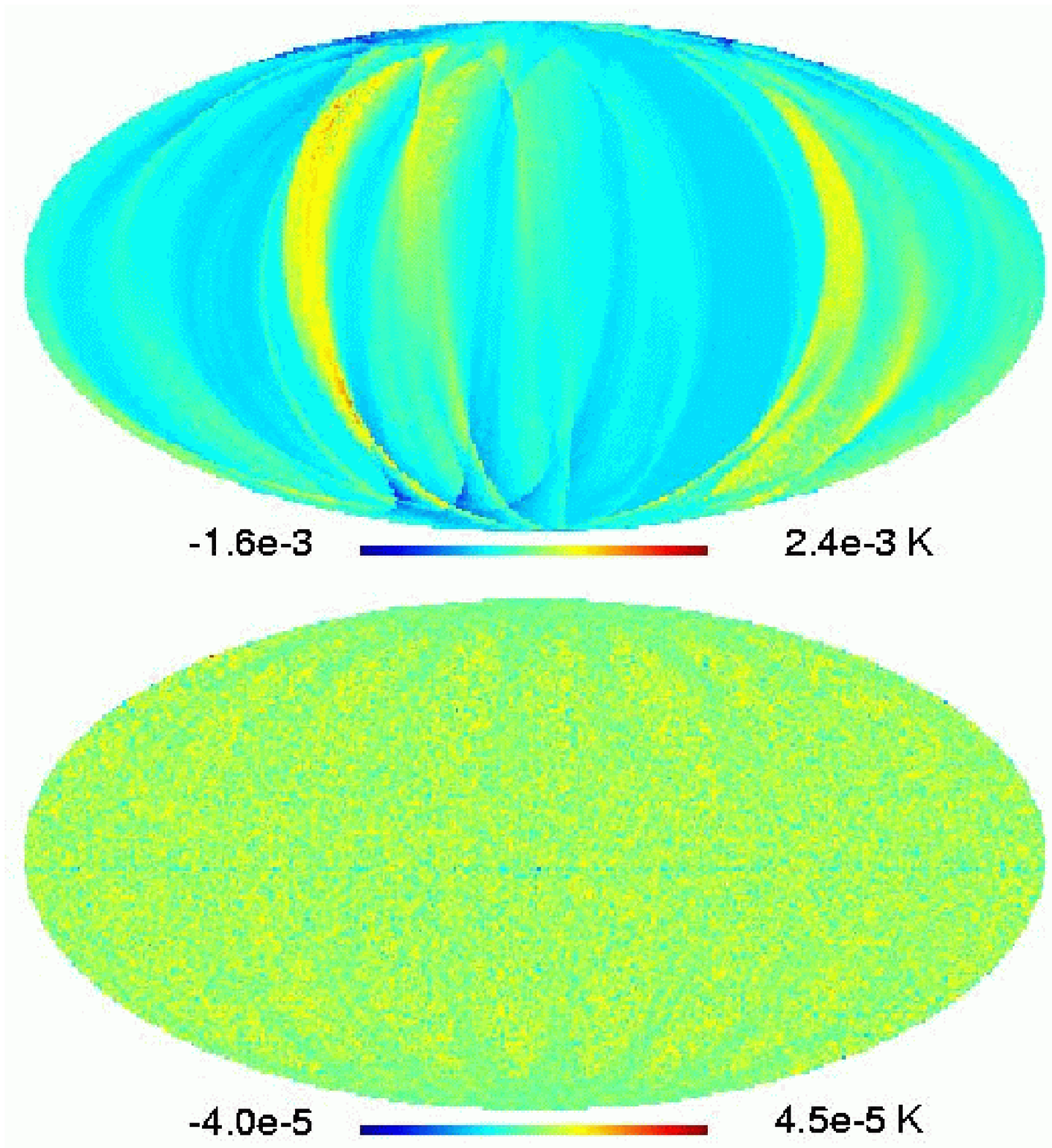}}
{\includegraphics[scale=0.4,clip=true]{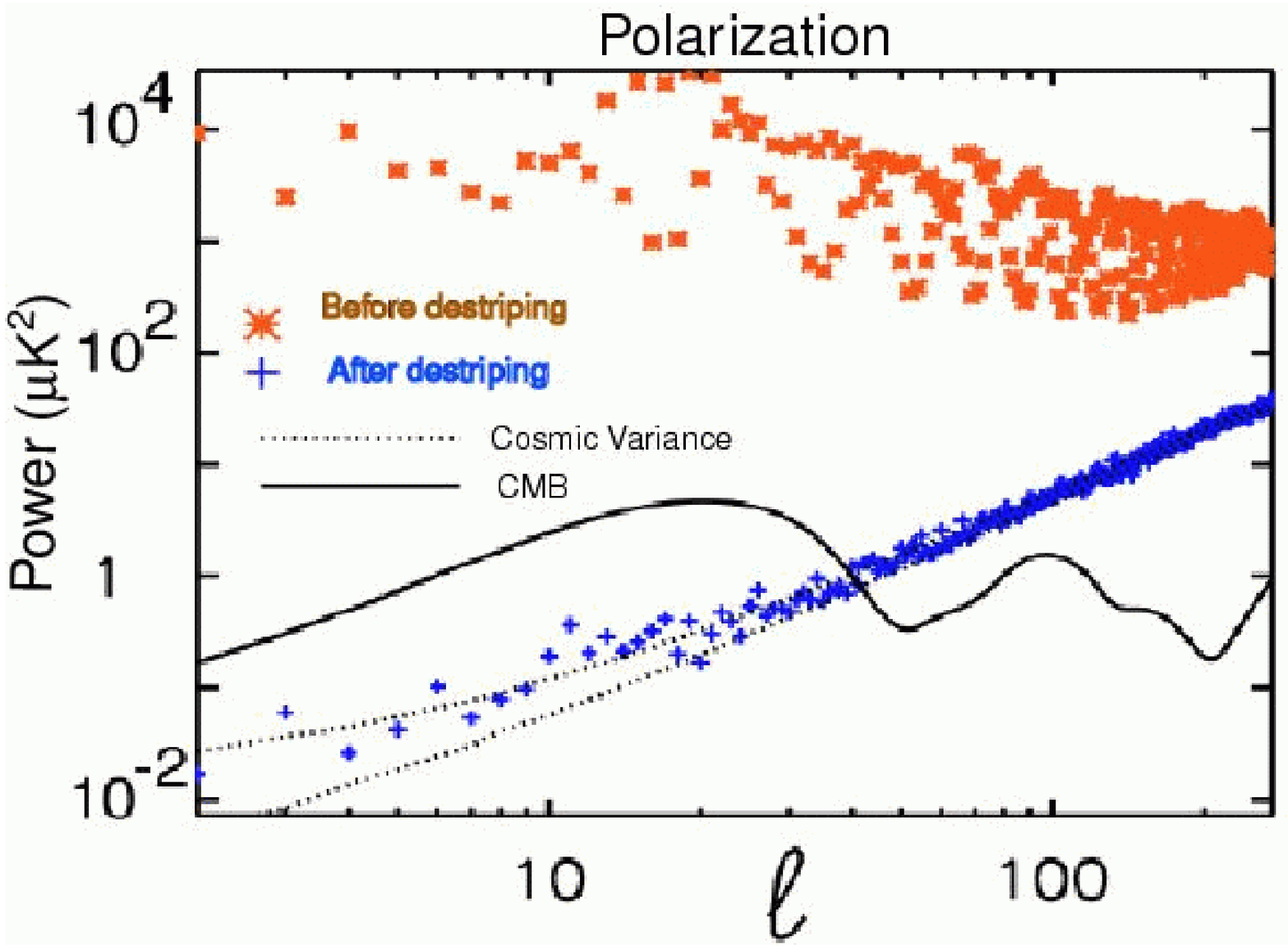}} 
    \caption{$Q$ residual maps before (upper left panel) and after
      (lower left panel) destriping. \newline The $E_l$ coefficients
      of the residual maps before
    and after destriping (right panel)}
    \label{fig:2}  
\end{center}
    \end{figure} 
    The quality of the "destriping" can be seen on figure \ref{fig:2}.
For further details, see \cite{revenu99}.
\section{Some useful definitions} 
\begin{figure}[!h]
\begin{center}
{\includegraphics[scale=0.4,clip=true]{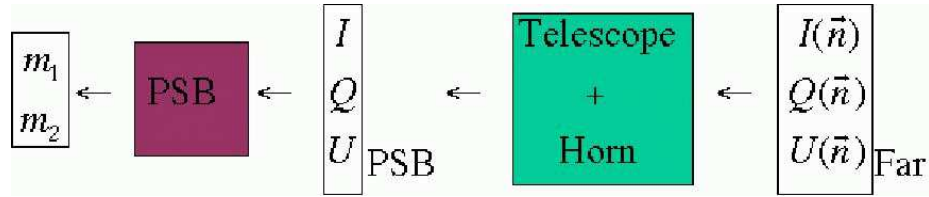}}
\caption{The path of a radiation entering a PSB\label{fig:2b}} 
\end{center}
\end{figure}

Let us consider a ``Polarization Sensitive Bolometer (PSB)''. The path 
followed by the incoming radiation is illustrated in figure \ref{fig:2b}.
The far field Stokes parameters  $(I, Q, U)_\mathrm{Far}(\vec{n})$
are integrated by going through the telescope and the horn to give the 
Stokes parameters on the PSB, $(I,Q,U)_\mathrm{PSB}$:
\beq
\begin{array}{lll}
I_\mathrm{PSB} &=& \int \left(a^I_I(\vec{n})\ I_\mathrm{Far}+
                        a^Q_I(\vec{n})\ Q_\mathrm{Far}+
                        a^U_I(\vec{n})\ U_\mathrm{Far}\right)\, d\vec{n}\\
Q_\mathrm{PSB} &=& \int \left(a^I_Q(\vec{n})\ I_\mathrm{Far}+
                        a^Q_Q(\vec{n})\ Q_\mathrm{Far}+
                        a^U_Q(\vec{n})\ U_\mathrm{Far}\right)\, d\vec{n}\\
U_\mathrm{PSB} &=& \int \left(a^I_U(\vec{n})\ I_\mathrm{Far}+
                        a^Q_U(\vec{n})\ Q_\mathrm{Far}+
                        a^U_U(\vec{n})\ U_\mathrm{Far}\right)\, d\vec{n}\\
\end{array}\label{eq:psb1}
\eeq
In principle the 9 real functions $a^I_I ...$ are needed to characterize the
beam completely. For a PSB at the center of a perfect
instrument, only  $a^I_I$ , $a^Q_Q$ and $a^U_U$ are present. The
signals measured by the two detectors of the PSB are given by: 
\beq
\begin{array}{lll}
m_1 &=& \frac{g_1}{2}\left((1+\epsilon)I_\mathrm{PSB} + 
         (1-\epsilon)(Q_\mathrm{PSB}\,\cos 2\alpha_1 +
                      U_\mathrm{PSB}\,\sin 2\alpha_1\right)\\
m_2 &=& \frac{g_2}{2}\left((1+\epsilon)I_\mathrm{PSB} - 
         (1-\epsilon)(Q_\mathrm{PSB}\,\cos 2\alpha_2 +
                      U_\mathrm{PSB}\,\sin 2\alpha_2\right)
\end{array}\label{eq:psb2}
\eeq
where $\epsilon$ is the rate of cross polarization leakage: if the
incoming radiation is not polarized, $\epsilon$ is the ratio of the
transmitted intensity polarized in the wrong direction to that 
  polarized in the right direction. The angles 
$\alpha_1$ ($\alpha_2$) are the angle between the polarised sensitive
direction 1 (2) and the $x$ ($y$) axis of the local reference
system. Ideally these two directions are exactly orthogonal and one
can choose the local reference frame so that  
$\alpha_1 = \alpha_2 = 0$ to remove the contribution of
$U_\mathrm{PSB}$, and make the three coefficients $a^I_U,a^Q_U$ and $a^U_U$ 
irrelevant. 
\\
The gain factors $g_1$ and $g_2$ are in general different. 

\section{Uncertainties on polarization leakage and polarizers
  orientations}
Uncertainties on the characteristics of the polarimeters will generate 
systematic errors. We focus here on two specific examples:
\begin{enumerate}
\item {\bf The cross-polarization leakage} $\epsilon$  in equation
  (\ref{eq:psb2}) is only known up to some uncertainty.
\item {\bf The polarimeter orientation in the sky} The angles
  $\alpha_1$ and $\alpha_2$ in equation (\ref{eq:psb2}) as well as the 
  relative orientation of the two PSB's necessary to measure the 3
  Stokes parameters are not exactly known either. 
\end{enumerate}
We have evaluated these effects as follows\\ 
i) 
Assume a ``theoretical setup'' of polarimeters with  given rates 
of cross-polarization leakage and given orientations in the sky.\\
ii) Assume that, due to imperfections in building the instrument, the
actual set up is different. An ``actual setup'' is built by adding random 
errors to the cross-polarization leakage and to the polarimeter
orientations.\\ iii) Observe a set of Stokes parameters $I,Q,U$ with the
``actual setup''\\
iv) Reconstruct the Stokes parameters using the ``theoretical
setup''\\
v) Compare the original and reconstructed Stokes parameters for a
random sequence of ``actual setups''.\\
The results are displayed in tables \ref{table:1} and \ref{table:2} ,
obtained with 4 polarimeters, $T=2.73$~K, $Q = U =
1\mu$K. Note that a known rate $\epsilon$ of cross-polarization 
  does not contribute to the systematic error but increases the statistical
  uncertainty by a factor $1/\sqrt{1-\epsilon}$
\begin{table}[!h]
\begin{center}
\begin{tabular}{ccc}
\hline
RMS error on & Relative & average error\\
leakage rates& RMS error on & on reconstructed\\
             & reconstructed polarization & polarization direction\\
\hline
0.01 & 1.4\% & 0.2$^\circ$\\
0.05 & 7\%   & 2$^\circ$\\
0.1  & 15\%  & 4.5$^\circ$\\
0.2  & 35\%  & 8$^\circ$\\
\hline
\end{tabular}
\caption{Errors due to uncertainties on the rates of cross-polarization 
  leakage}
\label{table:1}
\end{center}
\end{table}
\begin{table}[!h]
\begin{center}
\begin{tabular}{ccc}
\hline
RMS error on & Relative & average error\\
polarimeters orientations& RMS error on & on reconstructed\\
             & reconstructed polarization & polarization direction\\
\hline
0.1$^\circ$ & 0.2\% & 0.1$^\circ$\\
0.5$^\circ$ &   1\% & 0.5$^\circ$\\
  1$^\circ$ &   2\% & 0.9$^\circ$\\
  2$^\circ$ &   5\% &   2$^\circ$\\
  5$^\circ$ &  12\% &   5$^\circ$\\
 10$^\circ$ &  24\% &  10$^\circ$\\
\hline
\end{tabular}
\caption{Errors due the uncertainties on polarizer orientations}
\label{table:2}
\end{center}
\end{table}

In order to check that the above uncertainties do not impact on our
ability to measure polarized power spectra, we can test the effect of
an imperfect knowledge of polarimetric calibration parameters in yet
another way: a sky map simulated from a set of $C_l, E_l$ and $B_l$
spectra is observed with the ``actual setup''. The map is then
destriped as described above and reconstructed using the ``theoretical setup''.
\begin{figure}[!h]  
\begin{center}
\caption{The spectra of relative differences between input and output
  $E_l$  spectra}
\includegraphics[scale=.45]{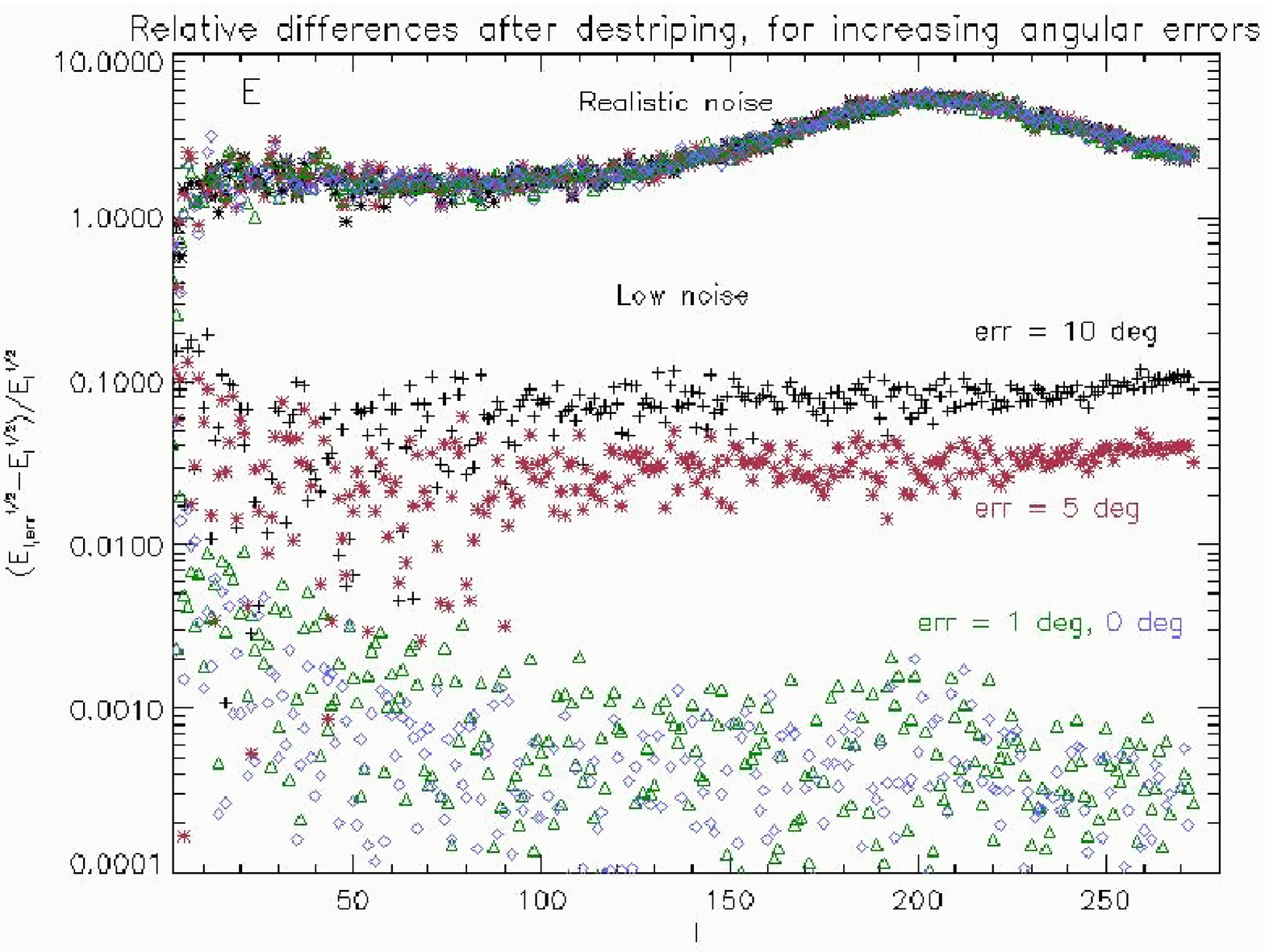}
    \label{fig:3}  
\end{center}
    \end{figure} 
 Finally the  $C_l, E_l$
and $B_l$ spectra of the reconstructed maps are computed and compared
with the inputs. Figure \ref{fig:3}  shows the result of this
comparison on the $E_l$ 
spectrum for random orientation errors of order 0 to 10 degrees. With 
a realistic noise the effects cannot be seen, therefore, the points
labelled ``low noise'' have been evaluated  with a noise divided by
$10^4$. With a $1^\circ$ error, the relative systematic errors on the $E_l$
spectrum remains below 1\% (actually closer to 0.1\%). The same results apply 
to the $B_l$ spectrum.

\section{The crucial problem: Signal differences}
As the $Q$ and $U$ parameters are computed by differences between
detector outputs, any  mismatch between the characteristics of the
detectors (gain, beam, pointing ...) induces a fake
polarization signal. 

\subsection{Relative photometric calibration between polarimeters
  (cross-calibration)}
The gains $g_1$ and $g_2$ in equation (\ref{eq:psb2}) are in general
different. This mismatch should be evaluated and corrected for by
cross-calibration. 
A residual cross-calibration error $\Delta g$ will result in a  
spurious polarization signal $\sqrt{Q^2+U^2} \sim \Delta g \times I$.
The $Q$ and $U$ fluctuations induced in this way are strongly
correlated to the temperature fluctuations.
For the CMB, a 1\% calibration error typically induces a 10\%-30\%
systematic error on the polarization fluctuation.
A constant calibration mismatch is easily detected and corrected
for. The trouble comes from gain variations with time. The time scale
for cross calibration have to be very carefully chosen. This question is 
currently under investigation on the polarized data from the Archeops
 flight in January 2000. Archeops is a balloon experiment to observe
 CMB fluctuations with an angular resolution around 10'. It involves
 21 bolometers at 143, 217, 353, and 545  GHz. The 6 bolometers at
 353  GHz are polarized and arranged in 3 Ortho Mode Transducers. After a
 technical flight in 1999, the first scientific flight occurred in
 January 2000 from the ESRANGE base at Kiruna in Sweden, and the data
 are currently being analyzed. Two more flights
 are planed in December 2001 and January 2002. (see
 Ref. \cite{archeops2001short} for more details).

\subsection{Pointing and/or beam shape mismatch}
The two orthogonal detectors necessary to obtain the $Q$ (or $U$)
parameter from the difference of their outputs should look at the same
area of the sky. However, this will in general not be true. 
Ortho Mode Transducers (OMT) or Polarized Sensitive Bolometers (PSB) 
(see the contributions of the SPORT and BOOMERANG teams to this
workshop) are a nearly perfect answer to this problem as the
polarization signal is the difference between the outputs of two detectors
sitting behind the same feed. However, even in this case a
pointing mismatch can occur if the time constants  of the 2
detectors are not the same. 
\begin{figure}[!h]  
\begin{center}
\includegraphics[scale=0.5]{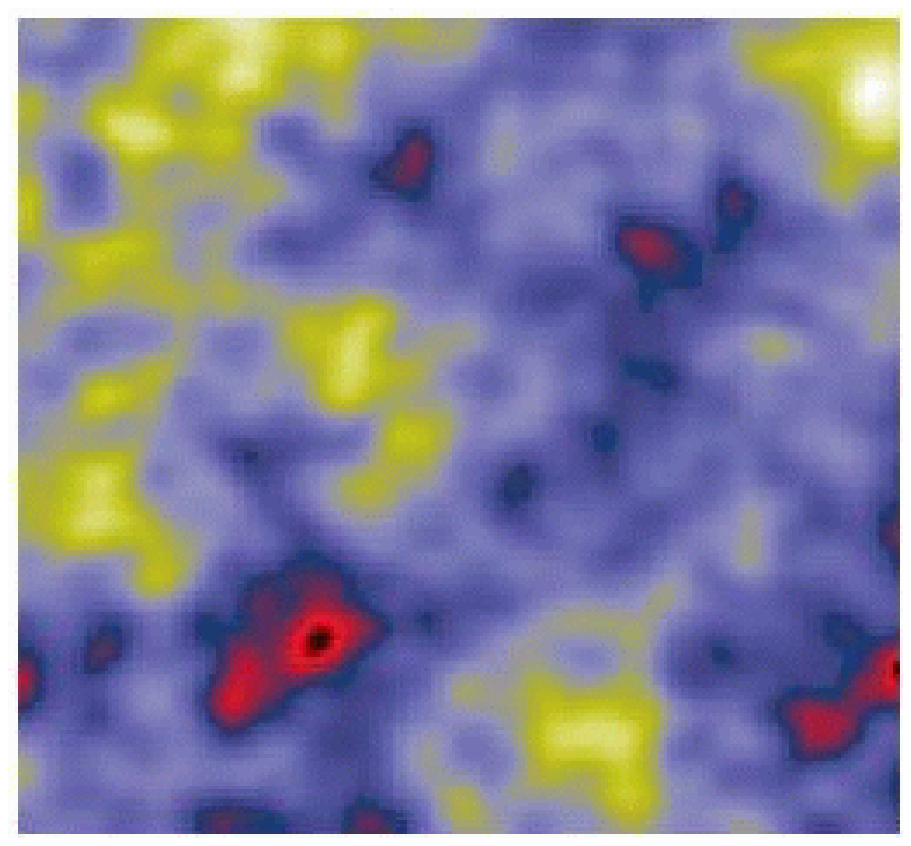}\hspace{.5cm}
\includegraphics[scale=0.5]{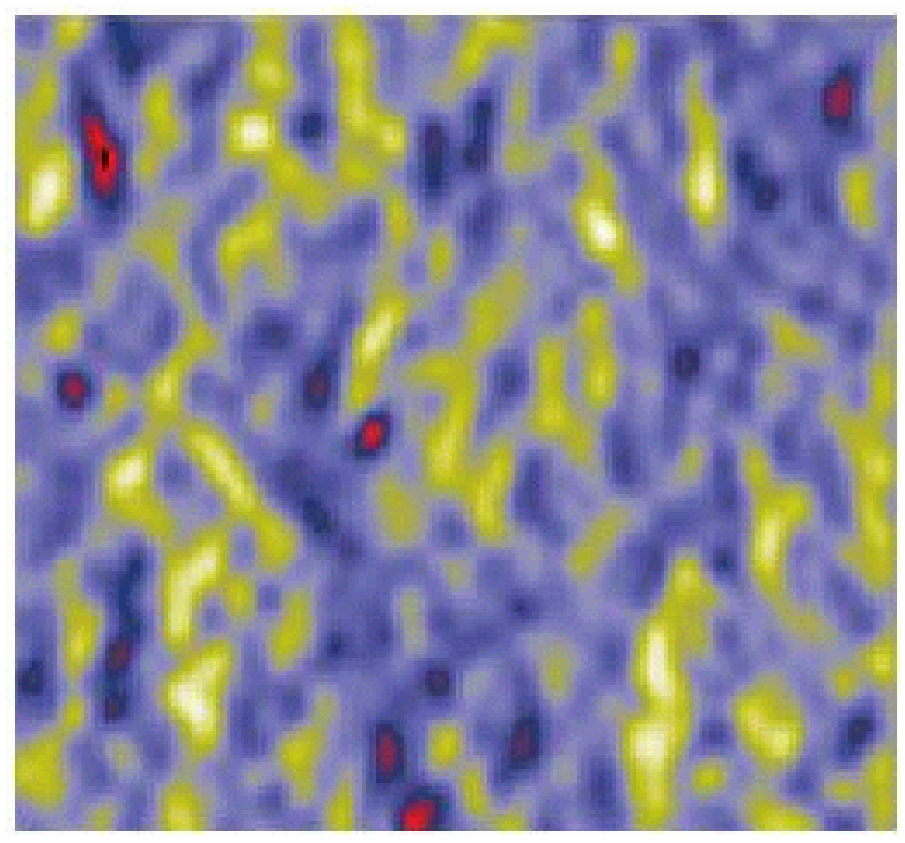}\hspace{.5cm}
\includegraphics[scale=0.5]{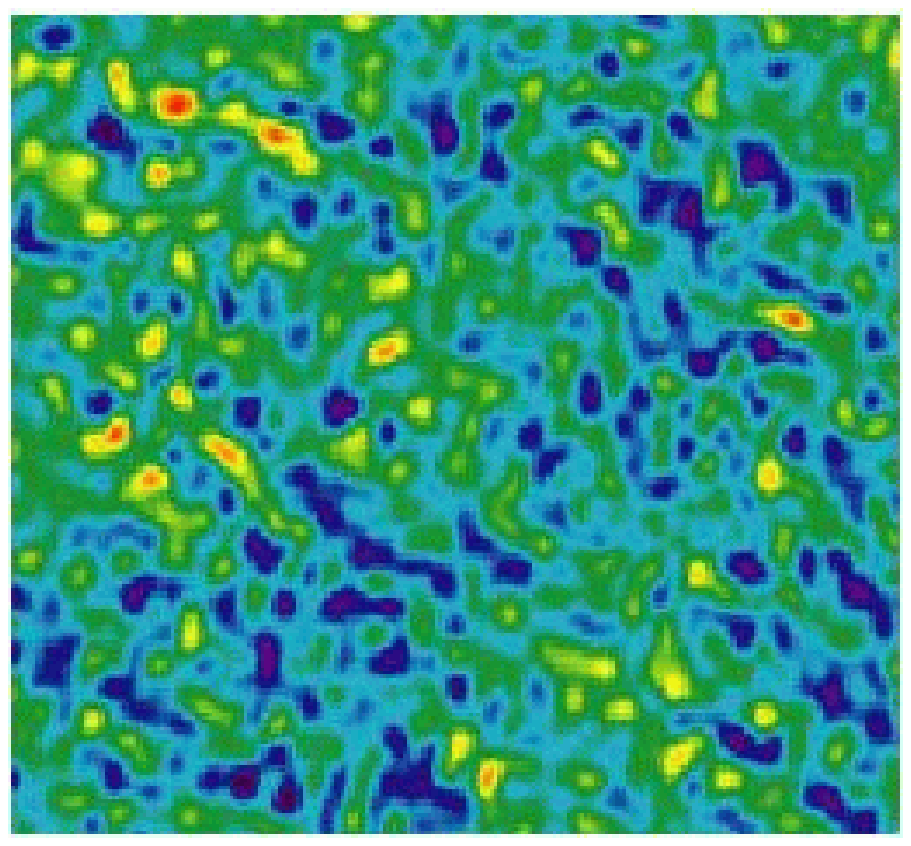}
\caption{The input temperature map (left) and the output $Q$ map with a 
  pointing mismatch of 0.5' (center) and with  the two
  mismatched beams shown in figure \ref{fig:5} (right)}
\label{fig:4}  
\end{center}
    \end{figure} 
\begin{figure}[!h]  
\begin{center}
\includegraphics[scale=0.4]{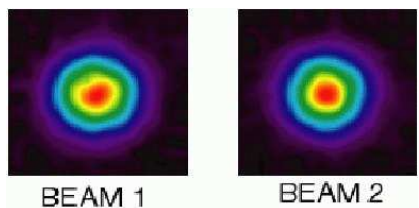}
 \caption{The two mismatched beams leading to the $Q$ map of the right 
   panel of figure \ref{fig:4}}
  \label{fig:5}  
\end{center}
    \end{figure} 
This is illustrated in figure \ref{fig:4}:
in the Planck mission, a 1 ms difference in the time constants of the two
orthogonal detectors induces a 0.5' pointing mismatch in the scanning
direction. A beam shape mismatch can arise for the same reason and also
because the two polarimeters are not oriented in the same way with
respect to the horn and the telescope.

In figure \ref{fig:4}, a  $4.1^{\circ}\times 4.1^{\circ}$ temperature map with a dispersion
$T_{RMS}=3.4\,10^{-5}$ and  zero
polarization (left panel), is observed with two orthogonal polarized
detectors. The center panel shows the $Q$ map induced by a pointing
mismatch of 0.5' between the two beams, otherwise identical (Gaussian and 7.5' wide). The output 
map develops $Q$ fluctuations with  $Q_{RMS}=1.2\,10^{-6}$, correlated 
to the temperature signal. This level is large and only slightly
smaller than the expected polarization level of CMB fluctuations. 
Note that for beams and beam mismatches small compared to the typical CMB
structures, the effect grows linearly with the distance between the
two beams.\\  
The right panel shows 
the $Q$ map generated by observing the same input temperature map with
the two mismatched beams shown in figure \ref{fig:5}. The two beams
differ by 2.5\% on 1/3 beam size scales. In this case the Q
fluctuation has  $Q_{RMS}=3\,10^{-7}$.

\section{Optimized polarimeter configurations}
Because of the low signal to noise ratio, an elliptic error box in the
$Q, U$ plane can induce a bias on the polarimeter direction. An
elliptic error box means unequal and/or correlated errors on $Q$ and
$U$. 

It can be shown \cite{couchot98} that a circular error box is obtained if\\ 
\begin{figure}[!h]  
\begin{center}
 \caption{An optimized configuration of 4 polarized detectors
   involving two OMT's or PSB's at 45$^\circ$ from each other. In
   reality, the two pairs of polarimeters point in the same sky direction at
   different times.}
\includegraphics[scale=.5]{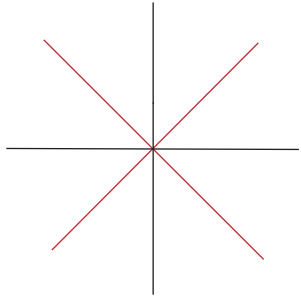}
  \label{fig:6}  
\end{center}
    \end{figure} 
i) the polarizer orientations are evenly distributed over
  180$^\circ$, as in figure \ref{fig:6} for 4 polarimeters\\
ii) the noise level should be as homogeneous as possible and
  uncorrelated among the   polarized detectors.

If these conditions are realized, one get as a bonus that the volume
of the error box in the $I, Q, U$ Stokes parameter is minimal.
Of course, this second condition will be the most difficult to realize 
in practice.

\section{Conclusions}
In an experiment such as Planck HFI, where polarization measurements
are made with detectors sensitive to the total polarized intensity in 
one direction, the main source of systematic
error in polarization measurements is the fact that $Q$ and $U$ Stokes
parameters are obtained from signal differences. This can be overcome and
even turned into an advantage if the systematics are common to both
detectors with the same size and therefore disappear in the difference.

OMT's and PSB's are a partial answer to this requirement as the two
detectors have nearly the same lobes and pointings. However the
electronic chains (and part of the optics for OMT's) are
different. Moreover one still has to combine the signals of two
different feeds to get the full polarized information. This latter
combination is less dangerous however, as it does not involve
intensity differences, and therefore will not generate polarisation
where there is none.

A rotating polarizing device in front of {\em one} bolometer in {\em
  one} feed and read by {\em one} electronic chain may provide a
solution to these difficulties, but one has to check that it does not
bring new systematics 
and it may be difficult to implement on satellite or balloon borne experiments.

  \bibliographystyle{aipproc}
\bibliography{cmb,mnemokap}
\end{document}